\documentstyle[preprint,prl,aps]{revtex}
\begin{document}
\preprint{\vbox{\hbox {September 1997} \hbox{IFP-737-UNC}  }}
%\twocolumn[\hsize\textwidth\columnwidth\hsize\csname@twocolumnfalse\endcsname
%\draft
\title{\bf Longevity and Highest-Energy Cosmic Rays}
\author{\bf Paul H. Frampton, Bettina Keszthelyi and Y. Jack Ng
}
\address{Department of Physics and Astronomy,}
\address{University of North Carolina, Chapel Hill, NC  27599-3255}
\maketitle
\begin{abstract}
It is proposed that the highest energy $\sim 10^{20}$eV cosmic
ray primaries are protons, decay products of a long-lived progenitor 
which has propagated from typically $\sim 100$Mpc. Such a scenario
can occur in {\it e.g.} $SU(15)$ grand unification and in some preon
models, but is more generic;
if true, these unusual cosmic rays provide a window into new physics.
\end{abstract}
%\vskip0pc]
\newpage

Important discoveries in particle physics were at one time dominated
by the study of cosmic rays\cite{gaisser}. Examples are the discovery of the positron\cite{positron},
the muon\cite{muon}, the pion\cite{pion}, and the first strange particles 
including the kaon\cite{kaon}.
In the last several decades, most of the important discoveries have been made
under the more controlled situation of accelerators. 
Nevertheless, the distinguished history for cosmic rays 
may be about to repeat, if the highest-energy cosmic 
rays reflect new physics.

The cosmic rays which exceed the GKZ cut-off\cite{GKZ}
at a few times $10^{19}$ eV are of particular interest, 
because for protons this cut-off which is based on pion 
photoproduction from the cosmic microwave background 
(CMB) seems very well founded. The derivation is as follows:
using for Boltzmann's constant $k_B = 8.62 \times 10^{-5}$eV/$^{o}$K 
and taking the temperature of the CMB as $3^{o}$K gives 
an average photon energy $\epsilon = 8\times 10^{-4}$eV. In the 
CMB frame a collision $p(p_1) + \gamma(p_2) \rightarrow 
\Delta \rightarrow N\pi$ has $p_1 = (E, 0, 0, k), 
p_2 = (\epsilon, 0, 0, -\epsilon)$ and squared center of 
mass energy $s = (p_1 + p_2)^2 = m_p^2 + 2(E + k)\epsilon
= m_{\Delta}^2$. For the relativistic case $E_p \simeq k$
and $E_p^{(resonance)} = (m_{\Delta}^2 - m_p^2)/4\epsilon =
2 \times 10^{20}$ eV. This is for the average energy: 
considering the more energetic CMB photons, 
the limit falls to a few $\times 10^{19}$eV. For protons above this energy, 
the pion photoproduction
from CMB will dominate beyond the mean free path. 
In the nonrelativistic case, when $E_p$ is not equal to $k$
one must keep all terms and find then 
$E_X + (E_X^2 - m_X^2)^{1/2} = 
(s_{threshold} - m_X^2)/2\epsilon$ where X is the 
primary and $s_{threshold}$ 
is the appropriate squared center of mass energy.
One could argue that for $E_p$ much larger than $E_p^{(GKZ)}$ there 
could be multiple scattering of the photoproduction type 
before the energy degrades to that
already observed for cosmic rays. But then there 
should be cosmic rays
of the higher energies without multiple scattering,
and it remains to be seen whether these are observed.

The mean free path ($\lambda$) of protons follows from the pion
photoproduction cross-section $\sigma = 200\mu$b (at the 
$\Delta(1236)$ resonance) and the density $\nu = 550$ 
photons/cm.$^3$ for the CMB. Thus,

\begin{equation}
\lambda = (200 \times 10^{-30} cm^2)^{-1} (550)^{-1} cm^3 = 
9\times 10^{24}cm. \simeq 3Mpc.  \label{mfpath}
\end{equation}
independent of the energy provided that $E_p >> m_p$.

This distance is an order of magnitude larger than our galactic 
halo ($\sim 100$kpc.) but is small compared to the Local 
Cluster ($\sim 100$Mpc.). It would suggest that the protons 
would need to 
originate within our galaxy, and hence be directed mainly from 
the galactic plane.
But the problem is that the maximal galactic 
fields are $\sim 3 \times 10^{-6}$G with 
coherence length $L\sim 300$pc. so a proton can typically be 
accelerated only to an energy:

\[
$$eBL \sim \left( \frac{4 \pi}{137} \right) ^{1/2} (3\times 10^{-6}G)(300pc.)
\simeq 10^{15} eV$$     
\]
several orders of magnitude too small. A further
problem is that such high energy $\sim 10^{20}$eV.
protons are hardly deflected 
by the interstellar magnetic fields
and hence should have a direction identifiable with
some source. 
To see this, note that for $E = 10^{11}$GeV, 
a proton of charge $1.6\times10^{-19}$ Coulombs in a magnetic
field $3\times10^{-6}$ gauss
has a minimum radius of curvature of $30$kpc. comparable
to the radius of the galaxy.
In fact, if anything, the eight
$> 10^{20}$eV. cosmic ray events
in hand are oriented along the exrtragalactic plane
and have no known correlation with any identifiable sources.
In short, these events are irresistible to a theorist.

These eight events are from
the AGASA\cite{AGASA}, Fly's Eye\cite{flyseye}, Haverah Park
\cite{haverahpark}, and Yukutsk\cite
{yukutsk} collaborations. The international Auger project\cite{auger}
would be able to find hundreds of such events
if it is constructed
and will likely shed light on the angular distribution
and on the presence of even higher energy
primaries. The existing events have shower properties 
and chemical composition
consistent with proton primaries.

Explanations offered for these extraordinary cosmic
rays have included protons originated from nearby
(but invisible otherwise) topological defects/monopolium
\cite{defects}, and magnetic monopoles\cite{KW}
that in the interstellar magnetic field can pick up
an amount of kinetic energy$(q_M/e) \sim 10^3$ times
higher than protons.

We investigate a different possibility. 
We hypothesize a particle 
$X$ with mass $M_X$ and lifetime $\tau_X$,
and whose cross-section with the CMB photons is below 6$\mu$b
so that the $\lambda$ in Eq.(\ref{mfpath}) is above 100Mpc.
The particle $X$ can be electrically neutral or charged. 
Suppose $X$ is like a heavy
quarkonium in QCD; then the linear size scales as $M^{-1}lnM$
and we expect a $500$TeV pseudoscalar to be several orders of magnitude
smaller than a proton and hence that
its cross-section $\sigma(\gamma X)$, which is $\sim(length)^2$
is correspondingly numerous orders
of magnitude smaller than $\sigma(\gamma p)=200\mu b$.
We assume that $X$ which obtained its kinetic energy as
the decay product of a GUT particle $G$
approximately at rest
decays into a proton within a few Mpc. of the Earth and
that this proton is the cosmic ray primary.

Let us consider $E_X = 2\times 10^{20}$ eV. and let $M_X$
be in GeV, $\tau_X$ in seconds and the distance $d$ be in Mpc. 
We assume $X$ is highly relativistic $E_X >> M_X$. Then
%\begin{equation}
\[ \frac{ \tau_{X} (sec) }{M_{X} (GeV) }  = 500 d (Mpc) \]
%\end{equation}
For a first example, take the neutron with $\tau_X \simeq 1000$
and $M_X\simeq 1$; this will travel 2Mpc.  It is an amusing
coincidence that this is close
to the proton mean free path, but it
means that the neutron does not
travel far enough to be the source we seek.

Let us revert to particle theory and ask for a neutral $X$ 
which will decay
very slowly into ordinary quarks.
Suppose, as an example, that there is a pseudoscalar
{\bf 27} of color $X^{\alpha\beta}_{\gamma\delta}$
coupling to four quarks by:

\[ \frac{1}{ M_{G}^3 } X^{\alpha\beta}_{\gamma\delta} \bar{q}_{\alpha} 
\gamma^5 q^{\gamma} \bar{q}_{\beta} q^{\delta} \]
with a suppression appropriate to a dimension-7 operator,
according to some GUT scale defect
with mass $M_G \simeq 2\times 10^{11}$GeV whose decay 
provides the kinetic energy $E_X >> M_X$. Because
$X$ is pseudoscalar, lower-dimension operators with gluons,
{\it e.g.} $XG^{\mu\nu}\tilde{G}_{\mu\nu}$ will be further suppressed.

The lifetime of $X$ may be roughly estimated
as:

\[ \tau_{X} \sim ( 10^{-23}sec.) \left( \frac{ M_{X} }{ 2 \times 10^{11} GeV }
\right)^{-6} \]
To fix parameters, let us take a distance scale 
$d=100Mpc.$ This fixes $M_X \simeq 500$TeV and 
$\tau_X \sim 3\times 10^{10} sec. \sim 1000$ years.

Is this picture natural from the point of view of the particle
theory? We need two scales: $M_G \simeq 2\times 10^{11}GeV$,
$M_X \simeq 500TeV$ and a pseudoscalar $X$ which is a {\bf 27} of
color, coupling to normal matter predominantly
by $d=7$ operators.

A specific example\cite{su15} occurs in $SU(15)$ where the GUT
scale can be $M_G = 2\times 10^{11}$GeV and the scale $M_X$
is identifiable with the scale called $M_A$ at which
$SU(12)_q$ breaks to $SU(6)_L\times SU(6)_R$.
For $X$ one can see by studying \cite{su15} that the 
{\bf 14,175} dimensional $SU(15)$ Higgs irreducible representation  
contains an appropriate candidate. 

Can the ancestor G particles at the GUT scale be uniformly distributed without
over-closing the universe? A crude estimate follows:

Take a spherical shell of radius $R$ and thickness $\Delta R$,
let the Earth radius be $R_{\oplus}$. For simplicity
assume the lifetime for G is the present age of the
universe $A = 10^{10}$y. The event rate in question is
$\sim 1/km^2/y$ at the Earth's surface so we have for $\nu_G/cm^3$
the number density of G particles:

\[ 1 \simeq \frac{1}{A} \left[ \nu_G \frac{4\pi }{3} (6R^2 \Delta R) \right]
\frac{\pi R_{\oplus}^2}{4 \pi R^2} \frac{1}{ 4\pi R_{\oplus}^2} \]
Putting $\Delta R = 2Mpc$, $A = 10^{10}y$, and $
M_G = 2\times 10^{11}GeV = 2\times 10^{-13}g$ gives
the mass density $\rho_G$:

\[ \rho_G = \nu_GM_G \sim 10^{-37} g/cm^3 \]
corresponding to a contribution $\Omega_G < 10^{-8}$ to
the closure density.
If the distribution of $G$ is not uniform but clustered in {\it e.g.} AGNs the $\Omega_G$ will presumably become
even smaller.
If we consider much larger $d >> 100Mpc$, the red-shift would
have to be taken into account in such calculations.
In all such cases, the required density of
GUT defects $G$ is consistent in the sense that it contributes 
negligibly
to the cosmic energy density.

This first example was merely an existence proof. The most unsatisfactory
feature is surely the color non-singlet nature of $X$.
It is difficult to believe such a colored state can
propagate freely even in intergalactic space because of color confinement.
A simple modification which avoids this difficulty
is to consider two such $X$ states: $X^{'}$ and $X^{''}$, both
{\it e.g.} color {\bf 27}-plets but non-identical.
The bound state $(X^{'}\bar{X}^{''}) = X$
can then be color singlet, stable with respect to $X'-\bar{X}''$
annihilation and unstable only by virtue of the decay of its constituents.

Three key properties of the $X$ particle 
in our scenario are (1)long lifetime, {\it e.g.} $\sim1000$y.
if dimension-7 operator(s) mediate
decay of $X$ to quarks, (2) small cross-section, below $6\mu$b., with CMB
photons, and (3) significant branching ratio for decay to proton(s).

What other models can exemplify this scenario?
One other example we have found is based on a slight
modification of the sark model\cite{sark1,sark2}.
The neutral sark baryon, or "nark", discussed in\cite{sark2}
as a candidate for dark matter merely needs
to be slightly unstable, rather than completely
stable, to play the role of our $X$ particle.
Firstly, 
the nark satisfies property (2) because of scaling arguments
(its mass is taken to be $\sim 100$GeV) similar to those used above.
If there is a unified theory of sarks, one expects sark number
to be violated and depending sensitively on the mass scale
of unification this could give a nark lifetime in the
desired range, property (1), as well as a sufficiently high 
branching ratio to protons, property (3), but without an explicit unified theory
this is speculation. More work on this idea is warranted.

Properties (2) and (3) are, in general, not unexpected: the most 
constraining requirement is property (1).
If there exists a model in which such longevity of $X$ 
occurs naturally it would be a compelling candidate
to be a correct description.

\bigskip
\bigskip
\bigskip

This work was supported in part by the U. S. Department of Energy under
Grant No. DE-FG05-85ER-40219.

\end{document}